# A Sequential Supervised Machine Learning Approach for Cyber Attack Detection in a Smart Grid System

Yasir Ali Farrukh
*Electrical and Computer Engineering*
*NED University of Engineering and Technology*
Karachi, Pakistan
yasirali.farrukh@gmail.com

Zeeshan Ahmad
*Pritzker School of Molecular Engineering*
*The University of Chicago*
Chicago, IL, USA
azeeshan@uchicago.edu

Irfan Khan
*Clean and Resilient Energy Systems Lab*
*Texas A&M University*
Galveston, TX, USA
irfankhan@tamu.edu

Rajvikram Madurai Elavarasan
*Clean and Resilient Energy Systems Lab*
*Texas A&M University*
Galveston, TX, USA
rajvikram787@gmail.com

*Abstract*—Modern smart grid systems are heavily dependent on Information and Communication Technology, and this dependency makes them prone to cyber-attacks. The occurrence of a cyber-attack has increased in recent years resulting in substantial damage to power systems. For a reliable and stable operation, cyber protection, control, and detection techniques are becoming essential. Automated detection of cyberattacks with high accuracy is a challenge. To address this, we propose a two-layer hierarchical machine learning model having an accuracy of 95.44 % to improve the detection of cyberattacks. The first layer of the model is used to distinguish between the two modes of operation – normal state or cyberattack. The second layer is used to classify the state into different types of cyberattacks. The layered approach provides an opportunity for the model to focus its training on the targeted task of the layer, resulting in improvement in model accuracy. To validate the effectiveness of the proposed model, we compared its performance against other recent cyber attack detection models proposed in the literature.

## I. INTRODUCTION

With the technological advancement in instrumentation, communication, networking, and control, the conventional power system has transformed into a smart grid system. The flow rate and size of sensed smart grid signals have drastically increased in recent years [1],[2]. This has transformed the traditional power system into an intelligent and autonomous smart grid system. Smart grids are now capable of state estimation, forecasting, controlling/predicting abnormalities, and even providing support to market agents. On the one hand, this advancement has proved quite beneficial. It has improved the overall efficiency and reliability of the system. At the same time, it has come across new technological challenges, such as bad data injection and false tripping and cyber-attacks. This happens due to the vulnerability of the store data on accessible data centers.

Due to the vulnerability of stored data, a hacker may corrupt the data as it is transmitted to the power system. This event is commonly known as a cyber-attack and may result in complete electrical blackouts [3], failure of government infrastructures, and breaches of national security secrets. Therefore, it is critical to identify and rectify any such cyber-attacks in smart grid systems.

Recently, a lot of attention has turned towards the automated identification of cyberattacks. Several techniques and methods involving supervised and unsupervised machine learning (ML) algorithms have been proposed for the detection of cyber-attacks. Such models are provided data related to power, voltage, and current parameters during normal operation and during cyberattacks, which is used to identify cyberattacks after training. A supervised ML algorithm based on Support Vector Machine was proposed in [4]. However, the proposed model was unable to distinguish between different types of transmission and generation anomalies with cyber attacks. Similarly, in Ref. [5], the authors proposed a method using two Euclidean distance-based techniques for detection. This method had an improved accuracy along with low computational complexity but lacked a diverse range of attack scenarios in training and testing. An overall comparison between several mostly used algorithms is presented in [6], which concludes that Random Forest Classifier (RFC) surpasses all other algorithms in terms of weighted accuracy.

In terms of unsupervised and deep ML algorithms, a recently proposed method using stacked Autoencoders is introduced in [7], with a classification accuracy of 93.7%. This model utilizes a self-adaptive cuckoo search algorithm for optimizing its parameter. However, the complexity of this model is high and requires a lot of time in training. In addition to this, a detailed overview of lately proposed approaches has been presented in Ref. [8], stating their problems and limitations. The major issue highlighted in the used algorithms is related to overfitting and the high computational complexity of the model.

Even with the development of highly accurate and precise algorithms, research work is still headed towards developing more rigorous and practically applicable methods. Therefore, the need to have models that are robust and pertinent is essential. Here we propose a two-layer hierarchical random forest model to detect cyber-attacks in a smart grid system that is robust and efficient in terms of applicability. The proposed model divides the problem into two sub-problems, the upper-level sub-problem, and the lower-level sub-problem. In the upper-level sub-problem, it distinguishes between a natural event or an attack event. If the distinguished data is a potential attack event, it is propagated to the lower-level sub-problem that works on identifying the specific attack classes. In turn, this approach results in better training of the sub-level models with respect to their specific event classes, leading to higher accuracy and lower computational complexity.

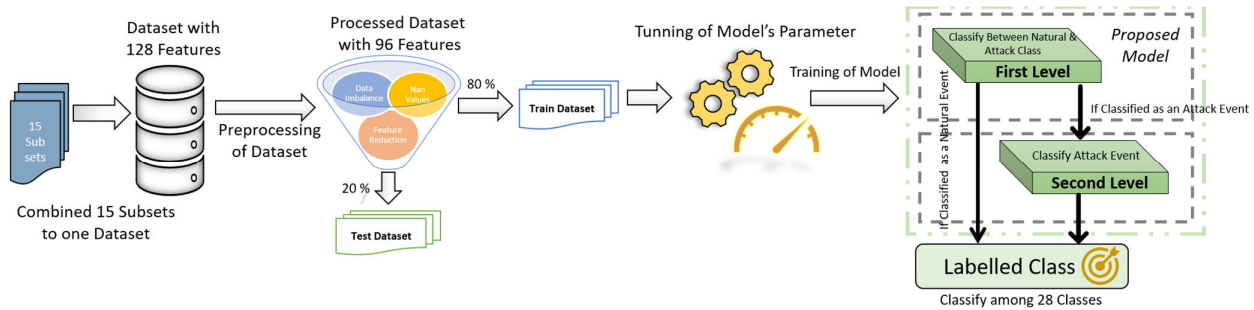

Fig. 1. Proposed methodology and model architecture. Dataset utilized is reduced to 96 features after preprocessing and is then divided into train and test set with a ratio of 8:2. The segregated test dataset is put aside and is not included in any testing or tunning of the model.

Before training the model, we perform a class balancing which is known to improve the accuracy of ML models. Our model achieves a classification accuracy of 95.44 %, which is higher than or similar to those obtained using other ML approaches.

The rest of the paper is organized as follows; Section II provides a detailed overview of the adopted methodology and proposes the model. Section III presents the results and their analysis, following section IV that concludes this research work.

## II. PROPOSED MODEL

Ref. [6] provides a comprehensive comparison of the performance of twelve commonly used ML algorithms for cyberattack detection. The random forest classifier (RFC) outperforms all other algorithms in classification accuracy. Therefore, RFC was used as the base model. Here, we optimize the RFC by exploring the layered hierarchical classification approach towards cyberattacks. Our choice for selecting RFC as a base classifier is due to its design robustness and stability. Moreover, it requires less computational cost as compared to deep learning techniques. Fig. 1 shows a schematic of the proposed model. The selection and training of the proposed model are explained in the further headings.

The dataset employed for this research work is available in 15 subsets. Instead of performing training and testing on each subset, an unabridged approach is adopted, and all subsets are combined to form one singular dataset. Before splitting the dataset into a train and test set, it is preprocessed by reducing its features from 128 to 96 features. Moreover, the class imbalance problem is also addressed in this part. After preprocessing, the dataset is divided with a ratio of 8:2 into train and test sets. The train set is utilized for training purposes while the test set is kept aside untouched for testing. The dataset used for the selection of parameters was a binary dataset instead of the multi-class dataset. Further details of the dataset are covered in sub-heading A.

### A. Overview of Data set Utilized

The dataset used for training our supervised model has been obtained from Adhikari et al. [9]. The power system dataset contains different parameters of four phasor measurement units (PMUs) related to the electric transmission system. There is a division of data set into three scenarios: binary, three class, and multi-class data set. The data set adopted for this experiment is multi-class data that comprises 78377 data points, having 37 different classes and 128 discrete features. The classes are distributed over the normal operation, fault condition, maintenance scenarios, data injection attack, attack on Relay setting, and remote tripping attacks. The ratio of data between these classes is 1:3:1: 2:9:2 [7]. Further detail of the dataset involving its parameter attribute, system structure, different classes, and approach can be found in Refs. [9, 10].

### B. Pre-processing of Data-set

Pre-processing is the initial step for solving problems related to data science. It is a method of converting raw data into a meaningful and understandable format. Raw data is mostly noisy, incomplete, and inconsistent. Therefore, in order to mitigate such problems, data should be preprocessed. Data preprocessing can be divided into the following three subparts:

Part 1: Handling of Missing or NaN Values

Identifying and coping up with missing values is essential for effective cyberattack identification research. If a researcher cannot handle missing values, it may cause the research to end up with an inaccurate deduction about the data. There are several approaches used for dealing with such problems. Among those techniques, the most popular ones are removing instances of Nan/missing values, replacing them with average value, majority value, or zeros. All of these methods are tested on a primary RFC having n_estimators set to 100. Moreover, this test run was carried out on a binary class dataset, and it is observed that replacing the NaN values with zero tends to produce the most effective results. The detailed outcome of each approach is shown in Table I.

TABLE I
HANDLING OF MISSING VALUE

| Approach | Accuracy Result |
| --- | --- |
| Mean | 94.51 % |
| Median | 94.02 % |
| Drop | 94.19 % |
| Replace with Zero | 94.54 % |

Part 2: Handling Imbalanced Classification

An imbalanced classification issue arises when the distribution among different classes of datasets is not uniform. This problem can lead to the poor performance of the classification model, especially for the minority class. Imbalance classification raises a challenge for training an effective model as most of the ML algorithms are designed for an equal number of datasets in each class. Many real-world problems are imbalanced in terms of datasets. If they are not handled properly, then the resulting outcome will be a biased trained model giving an edge to the majority class and overlooking the minority one [11]. The dataset is processed to overcome such a problem, and all the classes are balanced out before training the model. Several methods are used for handling such issues, which can be divided into two main approaches: data-driven and algorithm-driven [12].

For our research, the dataset used for training is highly imbalanced in terms of classes. Therefore to attain high model accuracy balancing out the dataset is needed. A data-driven approach is used for addressing this problem. Data resampling approaches are further divided into oversampling and undersampling [8].

Oversampling algorithms are adopted in order to achieve optimal efficiency while giving priority to the number of data instances. Moreover, a clearer picture of class imbalance in the dataset can be seen in Fig. 2

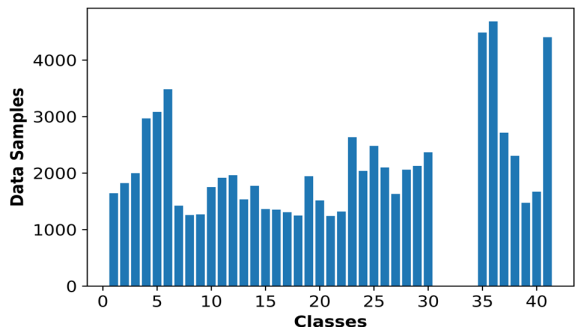

Fig. 2 Original population of classes in dataset, the population shows that classes are very imbalanced. Moreover, the classes between 30 and 35 do not exist in dataset, hence they have no values.

It can be apprehended from the figure that the number of data instances in each class is very inconsistent. The highest number of data instances is in class *36* with 4685 samples, while class *21* has the lowest number of data instances with 1242 samples only which are 3.7 times lesser than class *36*.

Oversampling techniques increase the number of data points rather than reducing them to uniformity, which is the case in undersampling. Moreover, undersampling approaches can only be used where the dataset is quite extensive, and loss in a small amount of data would not cause any significant change. However, in our case, we have a small amount of data. Therefore, undersampling was not an option. Regarding oversampling techniques, there are many approaches used for better results. Among them, the most popular are: Adaptive Synthetic (ADASYN) algorithm, Synthetic Minority Over-sampling Technique (SMOTE), Random Over Sampler (ROS), and Borderline SMOTE. All of these methods were applied to determine the most effective technique. The detailed results for each technique are presented in Table II.

TABLE II
METHODS FOR OVER SAMPLING

| Methods | Accuracy |
|---|---|
| None | 85.76 % |
| ADASYN | 93.89 % |
| Random Over Sampler | 96.23 % |
| SMOTE | 92.02 % |
| Borderline SMOTE | 93.72 % |

As per the results, ROS seems to have the best accuracy among all other approaches, but it tends to cause over-fitting for the model [12], [13]. In this method, minority classes are randomly replicated until the desired ratio of balancing is achieved. In order to avoid this problem, other modern techniques were considered. ADASYN and Borderline SMOTE algorithms have almost the same accuracy. The working of both the algorithm is also quite similar with just a little difference [14]. However, ADASYN generates synthetic data that is harder to learn, whereas SMOTE synthesizes data due to the interpolation of minority class datasets that are closely located [13]. These modern algorithms provide an adaptive approach for handling imbalance classification and helps in better training of the model. However, ADASYN has greater accuracy than the Borderline SMOTE, but it was not used. The reason for neglecting ADASYN is due to its density distribution criterion, which automatically calculates the number of samples to be synthesized for each minority class. While balancing the dataset, ADASYN exceeds the original maximum number of datasets in a particular class. Therefore, it was neglected to maintain the sensitivity of the model. The difference between SMOTE and Borderline SMOTE algorithm is that Borderline SMOTE is a variation of SMOTE algorithm. It only synthesizes data and the decision boundary present between the classes. In contrast, SMOTE algorithm generates synthetic data with its k-nearest neighbors of minority class [12].

A picture of datasets before and after the application of the Borderline SMOTE algorithm is shown in Fig.3. In addition to that, the balanced number of datasets in each class after oversampling is shown in Fig. 4. It can be seen that after oversampling of the dataset, all the classes are brought into unity in terms of data instances; that is, all classes now comprise 4685 data samples. Moreover, the difference of class distribution before and after oversampling of the dataset is exhibited in Fig. 4, which utilizes voltage and current magnitude of PMU 1 for visualization.

Part 3: Handling standardization of dataset

Data standardization is essential before implementing ML algorithms, as data standardization can significantly impact the outcome of the ML training model. Therefore, it is very crucial to have all the data on the same scale. There are many approaches available for data standardization. The methods tested for this particular experiment are described below. Among these methods, Standard Scaler (SS) acquired the highest accuracy of 95.2% on binary datasets. Therefore, it was utilized in this experiment.

Standard Scaler: transform all data features to the same magnitude, keeping mean 0 and variance 1. It does not involve any minimum and maximum value of the features, as shown in (1).

$$Standard\ Scaler = \frac{x_i - \bar{x}}{\sigma} \quad (1)$$

Mean Normalization: transform all data features such that the feature vector has one as Euclidian length. Scaling is done through different numbers for every data point as given in (2)

$$x' = \frac{x - \bar{x}}{\max(x) - \min(x)} \quad (2)$$

Min-Max Scaling: transform all the feature dataset to a scale between 1 and 0. It involves the computation of maximum and minimum values in the entire dataset, as given in (3)

$$x' = \frac{x - \min(x)}{\max(x) - \min(x)} \quad (3)$$

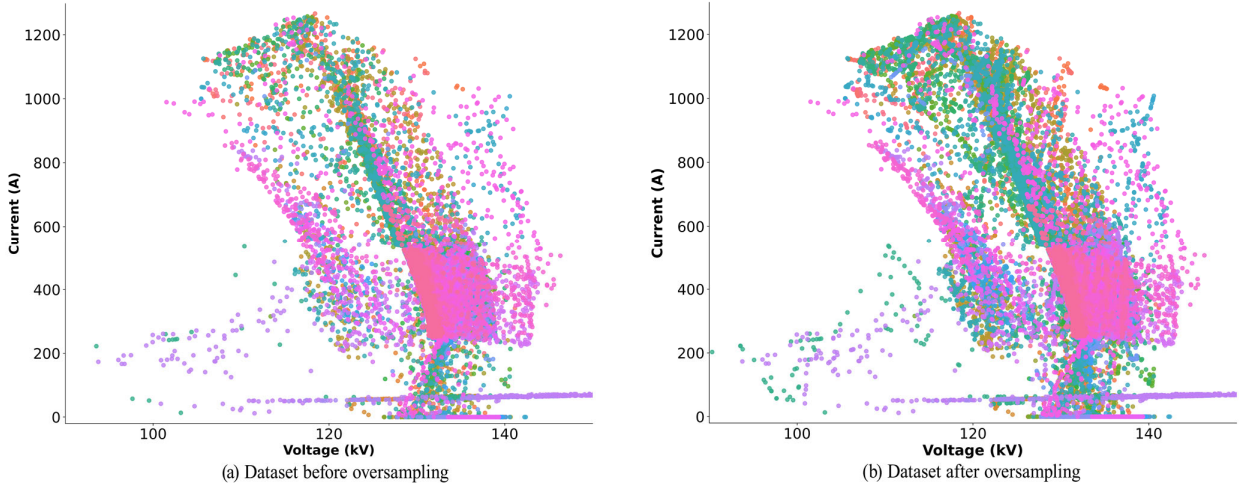

Fig.3. Scatter plot of the dataset, representing the distribution of classes with respect to two features, i.e., voltage and current. Different colors in the plot represent individual classes. (a) portrays composition of the dataset before oversampling by Borderline Smote algorithm and (b) portrays composition of the dataset after oversampling by Borderline Smote.

## C. Dimensionality Reduction

The number of the different features present in a dataset is known as the dimensionality of the dataset. As the number of data features increases, training a model becomes challenging. It will require more computational power and may lead to overfitting, resulting in performance degradation [10]. This issue is often termed as a curse of dimensionality. To mitigate such issue, high dimensionality statistics and different reduction techniques are used for data visualization. These methods are also applied in ML for optimizing the outcome of a model. The method used in this research for identifying the importance of each feature is through a Mean Decrease in Impurity (MDI).

The MDI is a measure of feature importance in evaluating a target variable. It calculates an average of total decrement in node impurity, weighted by the ratio of samples for each feature reaching that particular node in a separate decision tree. Thus, a higher MDI indicates the higher importance of that particular feature. The MDI index(G) is defined in (4)

$$G = \sum_{i=1}^{n_c} p_i(1 - p_i) = 1 - \sum_{i=1}^{n_c} p_i^2 \quad (4)$$

where $n_c$ is the number of classes in the target variable and $p_i$ is the ratio of this class.

The decrease in impurity $I$ is then defined in (5)

$$I = G_{parent} - P_{split1}G_{split1} - P_{split2}G_{split2} \quad (5)$$

where $P$ is the data proportion of each split that takes up the relative parent node.

MDI for the top 10 features of the dataset can be seen in Fig. 5. After analyzing the dataset, it is observed that the control panel log, relay log, snort log, and status flag of each PMU have very little or no importance in the detection of cyberattacks. Moreover, when it is analyzed with respect to domain knowledge, these features have no influence over the power system. Therefore, these features were dropped from the original dataset, and the dimensionality of the dataset was reduced to 96 features from 128 features. The effect of this reduction in feature can be seen in Table III.

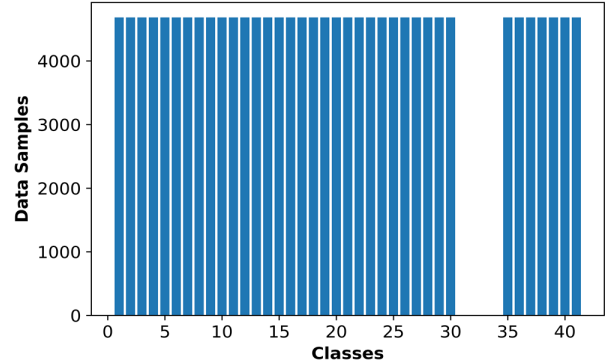

Fig. 4 Population of classes after oversampling with BoderlineSMOTE, each class is now balanced, having same number of data instances. classes between 30 and 35 do not exist in dataset, i.e. they are not sampled.

TABLE III
FEATURE REDUCTION

| Removal of Low MDI Features | Accuracy |
|---|---|
| Before | 94.51% |
| After | 94.59% |

After removing trivial features, the accuracy of the model increased. Keeping this approach in mind, another step was taken to further reduce dimensionality without losing any significant data. By using domain knowledge, voltage ($V$) and current ($I$) was converted to a single feature, Apparent power ($S$). The formula for $S$ is defined in (6)

$$S = V_{mag} * I_{mag} < \theta_V - \theta_I \quad (6)$$

where $\theta_V$ and $\theta_I$ represent the voltage and current angle.

Adopting this apporach, six features were reduced from each PMU, resulting in reducing 24 features altogether. After reduction, the dataset only contains 72 features. However, after training and testing the model again, only 0.02% in accuracy was increased. Since the increase in accuracy was not that substantial, therefore the dataset with 96 features was adopted as in the real-world problem having individual values of $I$, $V$, $\theta_V$ and $\theta_i$ is significant as they have high importance in terms of smart grid system operations.

*D. RFC Parameter tunning*

RFC is a well-known classification algorithm known for its robustness towards outliers. Moreover, it can handle noise comparatively better than other algorithm of its domain [6]. Like every other model, it is essential to tune the model as per the problem, to obtain effective results. Nebrase et al. in [6] have tested RFC with the same dataset, and the resulting outcome was 92%. For our particular research, the goal of parameter tunning was to improve the accuracy level in order to develop a sequential model capable of achieving higher accuracy results. There are several hyperparameters of RFC that can be adjusted. In this research work, only n_estimators, max_features, and criterion parameters of RFC were tested and tweaked for better accuracy. The details of the test can be found in Table IV.

. Table IV
RFC PARAMETER SELECTION

| Parameters | | Accuracy % |
|---|---|---|
| Max_Features | Sqrt | 94.54 % |
| | Log$_2$ | 94.87 % |
| Criterion | Gini | 95.08 % |
| | Entropy | 95.04 % |

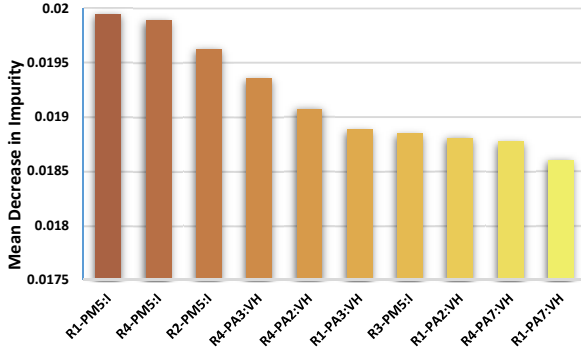

Fig. 5. Features having significant importance in distinguishing attack classes. Among 128 features, this figure represents the top 10 features in terms of mean decrease in impurity.

Different numbers of n_estimators were tested out to obtain the optimal value. The test range comprises 50 to 1000 n_estimaors. It was observed that the best result was obtained at 330 n_estimators. Therefore, the parameters used in this proposed model are Log$_2$ as max_features, Gini as the Criterion, and 330 n_estimators .

*E. Performance Metrics*

For the proposed model and training, evaluation criteria were set on the accuracy, recall, precision, and F1 score as shown in (7),(8),(9), and (10), respectively.

$$Accuracy = (TP + TN) / (TP + FP + FN + TN) \quad (7)$$

$$Precision = TP/(TP + FP) \quad (8)$$

$$Recall = TP/(TP + FN) \quad (9)$$

$$F1\ Score = 2TP/(2TP + FN + FP) \quad (10)$$

where *TP* and *TN* refer to true positive and true negative. Similarly, *FP* and *FN* refer to a false positive and false negative, respectively.

### III. RESULTS AND ANALYSIS

The core objective of this research work was to develop a sequential model having better accuracy and precision along with low computational cost. For achieving this goal, a bi-level model is proposed using RFC as a base classifier for the detection of intrusion attacks in smart grid systems. The model is divided into two layers. The first level sub-problem classifies between the natural events and attack events. Through this level, all-natural events are classified and filtered. This layer has an accuracy level of 99% in detecting a natural event. The reason for having such high accuracy is the better learning of the class boundary of two major classes rather than learning for all individual classes. Further, the classification of natural events in their specific classes is not part of this model. The intention for developing this model was to detect and classify intrusion attacks in smart grid systems. All events related to either fault, operation, or maintenance comes under the umbrella of natural events. If the upper-level classifies the data as an attack event, then it is passed onto the lower-level sub-problem, which classifies the data on the basis of 27 classes of attacks. The overall accuracy of the model is 95.44 %.

For training and testing purposes, the dataset of multi-class having 37 different classes is utilized. To avoid overfitting of the model, train and test sets were split with the ratio of 8:2 in the starting, and the test set was kept aside for the final model testing. The remaining train set was utilized for training both the layers of the model. However, original class markers present in the data were remapped as per the layer requirement. As for the first layer, classes 1-6,13,14, and 41 were marked as 1 (Natural Event), and all other classes 7-12, 15-30, and 35-40 were marked as 0 (Attack Class). For lower-level, class markers 1-6,13,14, and 41 were removed as this layer is trained for classifying attack events only. By doing this, the proposed model is trained robustly for the accurate identification of individual classes.

In order to justify this proposed model, it is essential to compare it with its baseline, which is the single-level model having all the same parameters and training environment. Moreover, a primary RFC was also tested on the same training environment but with no defined parameters that are having default parameters. The result of this comparison is shown in Fig.7

Other than the baseline, if the proposed model is compared with the other models proposed by researchers, it can be deduced that our proposed model has better accuracy in terms of classifying intrusion attacks on a smart grid system. The model proposed by Hink et al. [15] has an accuracy of less than 90%, and the model proposed by Keshk et al. [16] has an accuracy of 90.2%. In addition to these models, Defu et al. proposed a novel model in [10] using RFC as a base classifier and achieved a weighted accuracy of 93.91%. If we compare our model in terms of intrusion detection in a smart grid system with the model proposed in [10], our model clearly outperforms it with an accuracy of 95.44%.

It can be deduced from the research work that data preprocessing plays a crucial part in model performance. From class balancing of the dataset to feature reduction and standardizing of data, it helps improve the model training and enhance the model's predictive efficiency. By addressing the class imbalance problem, we provided a better learning environment to our proposed model, providing improved efficiency.

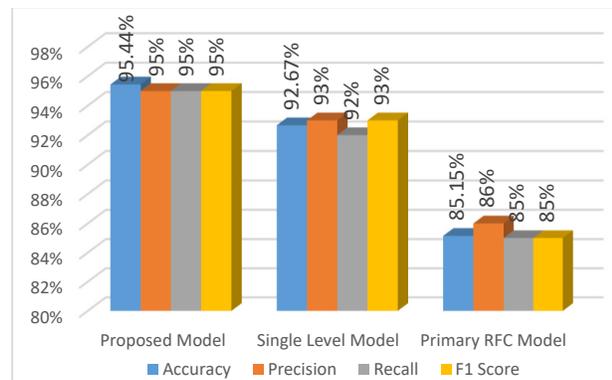

Fig. 7 Comparison betweem baseline models. All models are trained and tested in the same environment and same dataset. The difference between Single level model and the primary RFC model is of parameters. The primary model has default parameters, whereas single-layer model parameters are similar to the proposed model.

## IV. CONCLUSIONS

This study proposes a two-layered hierarchical approach with a baseline classifier to detect cyberattacks on a smart power system. We find that the two-layered traditional random forest algorithm performs better than deep learning algorithms. The limited attack data available makes it harder for deep learning approaches to learning the attack scenarios efficiently. Another issue in the currently used attack datasets is a class imbalance that results in model training heavily biased towards normal state instead of attack state. Tackling this issue before the training of the model through class balancing approaches can lead to improved performance of the current models. The performance results also reveal that feature reduction of the dataset can be quite useful, but it should be done considering the domain knowledge. The accuracy achieved by the proposed model is compared with the baseline models and found to outperform those for the detection of intrusion attacks in smart grid systems. Our study provides techniques to improve the accuracy of attack detection models while retaining the traditional ML algorithms with low computational costs.